\documentclass[slac_one]{revtex4}
\usepackage{graphicx}
\usepackage{fancyhdr}
\newcommand{\ETM}       {\rm E_T\hspace{-2.4ex}/\hspace{1.2ex}}
\newcommand{\MET}       {$\rm E_T\hspace{-2.4ex}/\hspace{1.2ex}~$}
\newcommand{\gl}        {\tilde{g}}
\newcommand{\sq}        {\tilde{q}}

\newcommand{\mGeVcc}    {\mathrm{GeV/c}^2}
\begin{document}

\title{SUSY Searches with Jets at CDF} 

%

\author{M. D'Onofrio}
\affiliation{Institut de F\'isica d'Altes Energies, E-08193 Bellaterra (Barcelona), Spain}
\begin{abstract}
We present the most recent results on searches for squarks and gluinos 
in $p\bar{p}$ collisions at $\sqrt{s}=1.96$ TeV, in events with large 
missing transverse energy, leptons and multiple jets in the final state, based 
on 1.8 to 2.8 fb$^{-1}$ of data collected by the CDF Run II detector at 
the Tevatron. No evidence of new physics is found and exclusion 
limits in several MSSM scenarios are extracted. 
\end{abstract}

\maketitle

\thispagestyle{fancy}

\section{INTRODUCTION} 

Supersymmetry (SUSY)~\cite{super} is regarded as one of the most
compelling theories to describe physics at arbitrarily high energies 
beyond the Standard Model (SM). In SUSY, a new spin-based symmetry 
turns a bosonic state into a fermionic state - and vice versa - 
postulating the existence of a superpartner for each of the known 
fundamental particles, with spin differing by 1/2 unit.  
The phenomenology is determined by the breaking mechanism of 
the symmetry. Several constraints are assumed to reduce the 
vast SUSY parameter space. In mSUGRA~\cite{msugra}, one of the 
most extensively studied models, symmetry breaking is achieved via 
gravitational interactions and only five parameters determine the 
low energy phenomenology from the scale of Grand 
Unification(GUT).

If R-parity~\cite{Rp} is conserved, SUSY particles have to be produced in pairs and
ultimately decay into the lightest supersymmetric particle (LSP), 
usually identified as the lightest neutralino $\tilde{\chi}^{0}_{1}$,
which constitutes a valid candidate for cold dark matter.

The production of squarks ($\sq$) and gluinos ($\gl$), superpartners of 
quarks and gluons, constitutes one of the most promising SUSY channels 
at the Tevatron because of the strong couplings involved. 
Their cascade decay will result in a final state consisting 
of several jets, leptons and missing transverse energy (\MET) 
coming from the neutralinos, which leave CDF undetected. 
In this document, we present the most recent results on searches for squarks and
gluinos in $p\bar{p}$ collisions based on data collected 
at the CDF experiment.

\section{INCLUSIVE SEARCH FOR SQUARK AND GLUINO PRODUCTION}
Squarks and gluinos are searched for in events with large \MET  
and multiple jets in the final state, using 2 fb$^{-1}$ of data~\cite{sqglpublic}. 
A R-parity conservation mSUGRA scenario is assumed, with the common 
soft trilinear SUSY breaking parameters $A_0 = 0$, the sign of the Higgsino mass 
term $\mu<$-1, and the ratio of the Higgs vacuum expectation values at the 
electroweak scale tan$\beta=5$.  

The gluino-squark mass plane is scanned via variations 
of the parameters $m_0$ (0-500 GeV/c$^{2}$) and $m_{1/2}$ (50-200 GeV/c$^{2}$), 
common scalar and gaugino mass at the GUT scale respectively.
The {\sc pythia}~\cite{pythia} Monte Carlo program is used to generate
samples for each mSUGRA point. Light-flavor squark masses are 
considered degenerate, while 2-to-2 processes involving stop ($\tilde{t}$)
and sbottom ($\tilde{b}$) production are excluded to avoid strong theoretical 
dependence on the mixing in the third generation.  
Squark/gluino production cross sections are normalized to next-to-leading 
order (NLO) predictions as estimated using {\sc prospino v.2}~\cite{prospino}.

Depending on the relative masses of $\sq$ and $\gl$, different event 
topologies are expected. If squarks are significantly lighter 
than gluinos, $\sq\sq$ production is enhanced. The squark tends to decay 
according to $\sq \rightarrow  q \tilde{\chi}^{0}_{1}$, and a dijet topology is
favoured, along with \MET due to the two neutralinos in the final state.
If gluinos are lighter than squarks, $\gl\gl$ process dominates.
Gluinos decay via $\gl \rightarrow q \bar{q} \tilde{\chi}^{0}_{1}$,
leading to topologies containing a large number of jets ($\geq 4$) and
moderate \MET. For $m_{\tilde{g}} \approx m_{\tilde{q}}$, a topology
with at least three jets in the final state is expected.
Three different analyses are carried out in parallel, requiring at least  
2, 3 or 4 jets in the final state, respectively. In each case, the transverse 
energy of the jets must be above 25 GeV, and one of the leading jet is
required to be central ($|\eta | < $1.1). Events are required to have a 
reconstructed primary vertex with $z$-position within 60 cm of the nominal 
interaction point and $\ETM>$70 GeV. In order to reject cosmics and beam-halo 
background, events are also required to have a tracking activity consistent with the 
energy measured in the calorimeter. 

SM background contributions are estimated using Monte Carlo simulation 
samples. QCD multijet processes, where the observed 
$\ETM$ comes from partially reconstructed jets in the final 
state, dominate. A minimum azimuthal distance between the jets and the
\MET is required to reduce this contribution. 
Lepton vetoes are applied to reject other sources of backgrounds such as 
Z and W production in association with jets, and top and diboson production. 
Monte Carlo predictions for SM processes are tested in
background-dominated regions, referred as control regions,
defined by reversing the selection requirements introduced
to suppress specific background processes. Good agreement is found
between data and SM predictions in all control samples
considered for each separate analysis. 

For each final state with different jet multiplicity, a dedicated study has been carried 
out to define the selection criteria that enhance the sensitivity
to the mSUGRA signals and further reject background contribution. 
The signal significance, S/$\sqrt{\rm B}$, with S denoting the signal 
and B the background number of events, is maximized. 
Jet transverse energies, $\ETM$ and $\rm H^{}_{T}$, the later defined as the sum of 
the transverse energy of the jets, are the variables employed in the 
optimization. Table I summarizes the thresholds applied on 
these variables in the different analyses. 

The number of observed and SM expected events corresponding to
a total integrated luminosity of 2.0 $\rm fb^{-1}$ are reported in 
Table II, separately for the three analyses. 
The total systematic uncertainty on the SM predictions varies 
between 31$\%$ and 35$\%$ as the jet multiplicity increases
and is dominated by the 3$\%$ uncertainty on the jet energy scale. 

\begin{figure*}[t]
\mbox{
\begin{minipage}{0.45\textwidth}
\centerline{\mbox{\includegraphics[width=85mm,height=78mm]{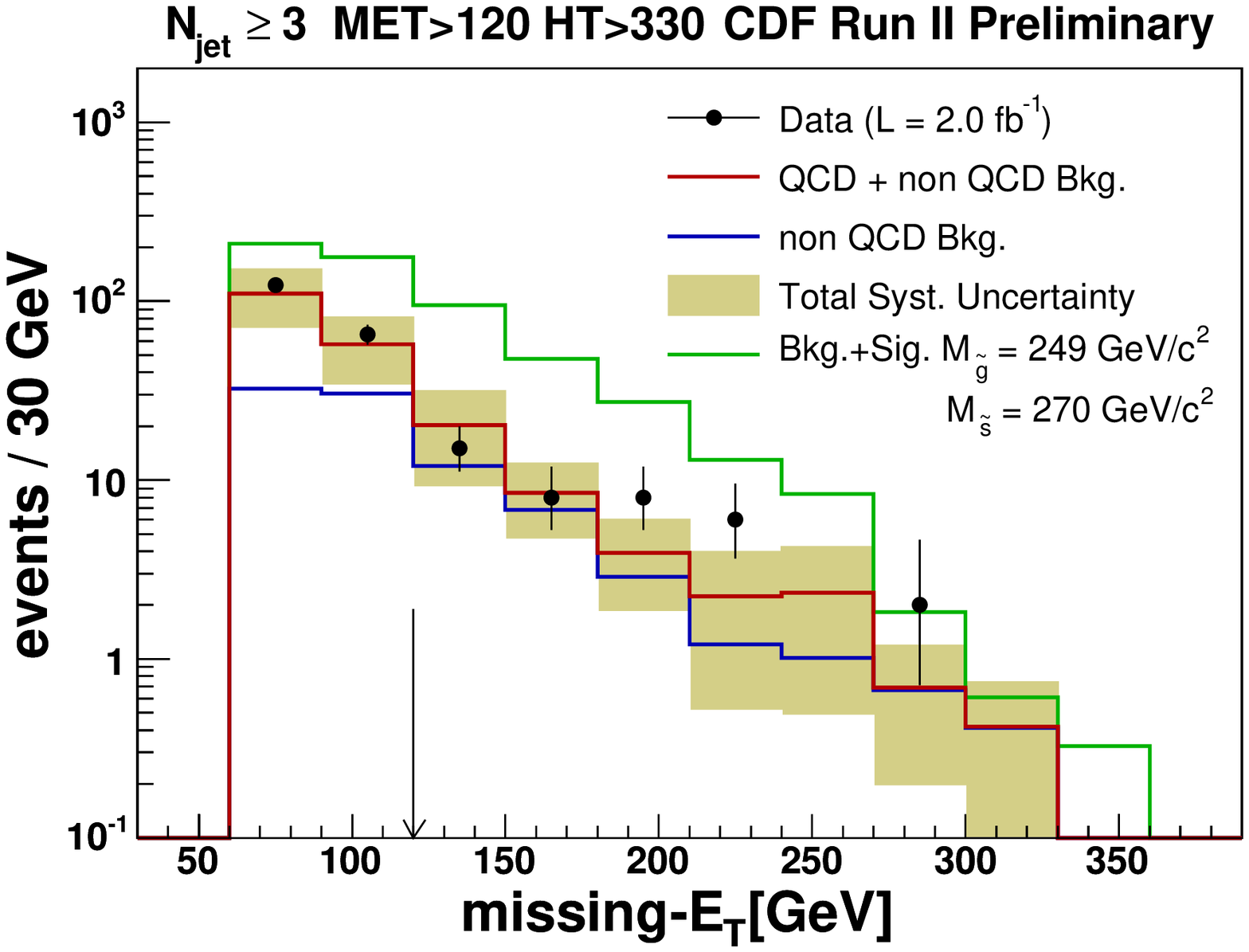}}}
\vspace{-0.02in}
\end{minipage}\hspace*{0.4in}
\begin{minipage}{0.45\textwidth}
\centerline{\mbox{\includegraphics[width=90mm]{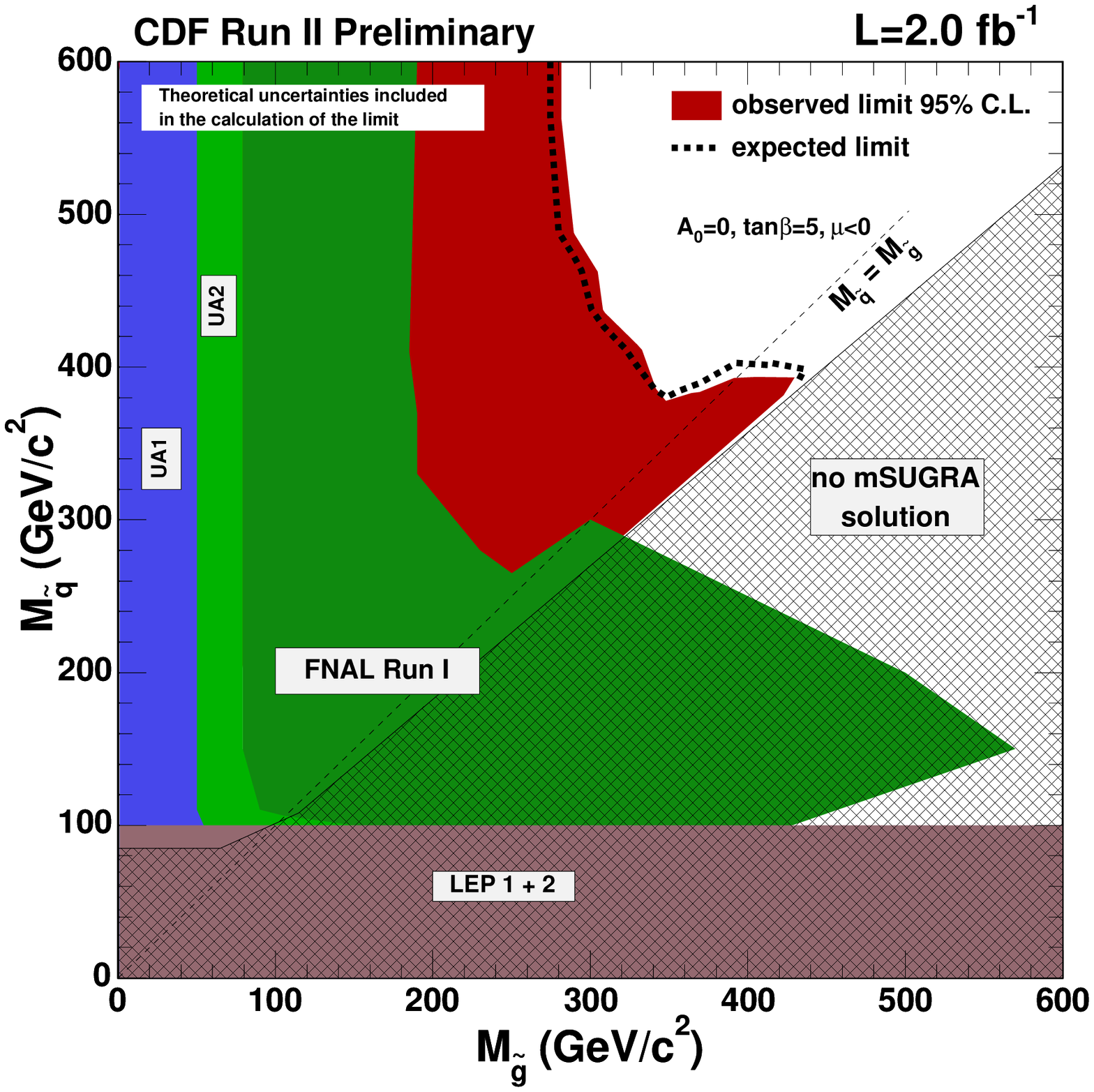}}}
\vspace{-0.01in}
\end{minipage}
}
\caption{{\bf LEFT:} $\ETM$ distribution for the 3-jets region. All selection criteria have 
been applied except the one on the variable that is represented. The arrow 
indicates where the cut on $\ETM$ is placed. {\bf RIGHT:} Squark-gluino mass plane: 
the dark-red region shows the area excluded at 95$\%$ C.L. by the current analysis 
with 2.0 fb$^{-1}$ of CDF Run II data. The dashed black line shows the expected 
limit. Previous exclusion limits are also reported.} \label{fig:sqgl}
\end{figure*}

Figure~\ref{fig:sqgl}(left) shows the measured $\ETM$ distribution for the
3-jet analysis case -- where the arrow indicates the final cut on $\ETM$ -- 
compared to SM predictions after all selection criteria have been applied.  
For illustrative purpose, the signal for one mSUGRA point is also shown.
No significant deviation from SM predictions is found. The results 
are translated into 95$\%$ C.L. upper limits on the cross section for squark 
and gluino production in different regions of the squark-gluino mass plane,
using a Bayesian approach and including statistical and systematic uncertainties.
Figure~\ref{fig:sqgl}(right) shows the excluded region in 
the squark-gluino mass plane. This search excludes masses up 
to $392 \,  ~\mGeVcc$ at 95$\%$ C.L. in the region where gluino and 
squark masses are similar and gluino masses up to $280 \, ~\mGeVcc$ for 
every squark mass.

\begin{table}[h!]
\mbox{
\begin{minipage}{0.45\textwidth}
\begin{tabular}{cccc}
\hline\noalign{\smallskip}
Variable [GeV] & 4-jet & 3-jet & 2-jet  \\
\noalign{\smallskip}\hline\noalign{\smallskip}
$\rm H^{}_{T}$ & 280 & 330 & 330 \\ 
\MET & 90 & 120 & 180 \\ 
$\rm E^{jet1}_{T}$ & 95 & 140 & 165 \\ 
$\rm E^{jet2}_{T}$ & 55 & 100 & 100 \\ 
$\rm E^{jet3}_{T}$ & 55 & 25  & -- \\ 
$\rm E^{jet4}_{T}$ & 25 & --  & -- \\ 
\noalign{\smallskip}\hline
\end{tabular}
\caption{Set of thresholds employed in the inclusive search for squark and gluino production.}
\label{tab:1}       
\vspace*{0.0cm}  
\end{minipage}\hspace*{0.4in}

\begin{minipage}{0.45\textwidth}
\begin{tabular}{lcc}
\hline\noalign{\smallskip}
Region & Data Evts  &  SM Exp. Evts \\
       &     &  (stat. $\oplus$ syst.) \\
\noalign{\smallskip}\hline\noalign{\smallskip}
4-jets    & 45   & 48 $\pm$ 17 \\ 
3-jets    & 38   & 37 $\pm$ 12  \\
2-jets    & 18   & 16 $\pm$  5  \\ \hline
\noalign{\smallskip}\hline
\end{tabular}
\caption{Observed number of data events for the three  
selection analyses in 2.0 fb$^{-1}$, compared with the 
expected events from SM processes. 
The quoted systematic uncertainty 
on the background include 6$\%$ uncertainty on the luminosity.}
\label{tab:2}       
\vspace*{-0.6cm}  
\end{minipage}}
\end{table}

\section{SEARCH FOR GLUINO-MEDIATED SBOTTOM PRODUCTION}

In MSSM, for high values of tan$\beta$, the large mass splitting in 
the third generation sector yields to low masses for the lightest sbottom state. 
If $m^{}_{\gl}$ is lower than $m^{}_{\sq}$~\footnote{where $\sq$ are first and second 
generation squarks, degenerate in mass.} but larger than $m^{}_{\tilde{b}}$, gluino pair 
production dominates. Assuming 100$\%$ branching ratio (BR) for  
$\gl \rightarrow \tilde{b} b$ with subsequent sbottom decay to a 
b-quark and the lightest neutralino ($\tilde{b} \rightarrow \tilde{\chi}^{0}_{1} b$), 
final state events with four b-jets and missing transverse energy are expected. 

This search is based on 1.8 fb$^{-1}$ of data~\cite{sbottomnew}. 
Signal samples are generated using {\sc pythia} ($\tilde{\chi}^{0}_{1}=60$GeV/c$^{2}$, 
$m^{}_{\sq}=500$ GeV/c$^{2}$), scanning gluino masses between 240 GeV/c$^{2}$ and 
400 GeV/c$^{2}$ and sbottom masses between 150 GeV/c$^{2}$ and 350 GeV/c$^{2}$.
The total NLO production cross section of $\gl \gl$ process is estimated using 
{\sc prospino}. 

Following a similar strategy to the one adopted in the inclusive $\sq/\gl$ search, 
requirements on tracking/calorimeter activities are applied to reject cosmics and 
beam-halo background. Events are required to have a reconstructed primary vertex 
with $z$-position within 60 cm of the nominal interaction point, 
$\ETM$ above 70 GeV, and at least two jets with transverse energy above 25 GeV. 
To enhance sensitivity to multi-b jets final state, at least one of the two 
leading jets is required to be originating from a b-quark. Such jets are identified 
using a tagging algorithm based on the reconstruction of secondary vertex within the 
jet cone. 

W and Z boson production in association with jets and top production 
represent the dominant sources of SM background and are estimated using Monte Carlo 
simulation. Background contributions from QCD multijet production and light flavor 
jets misidentified as b-jets ('mistag') are estimated from data. 
The prediction for SM processes is tested in control regions 
defined such that they result in background-dominated samples in which signal 
contribution is negligible. Good agreement between data and MC predictions is found 
in all control samples. 

\begin{figure*}[t]
\mbox{
\begin{minipage}{0.45\textwidth}
\centerline{\mbox{\includegraphics[width=90mm]{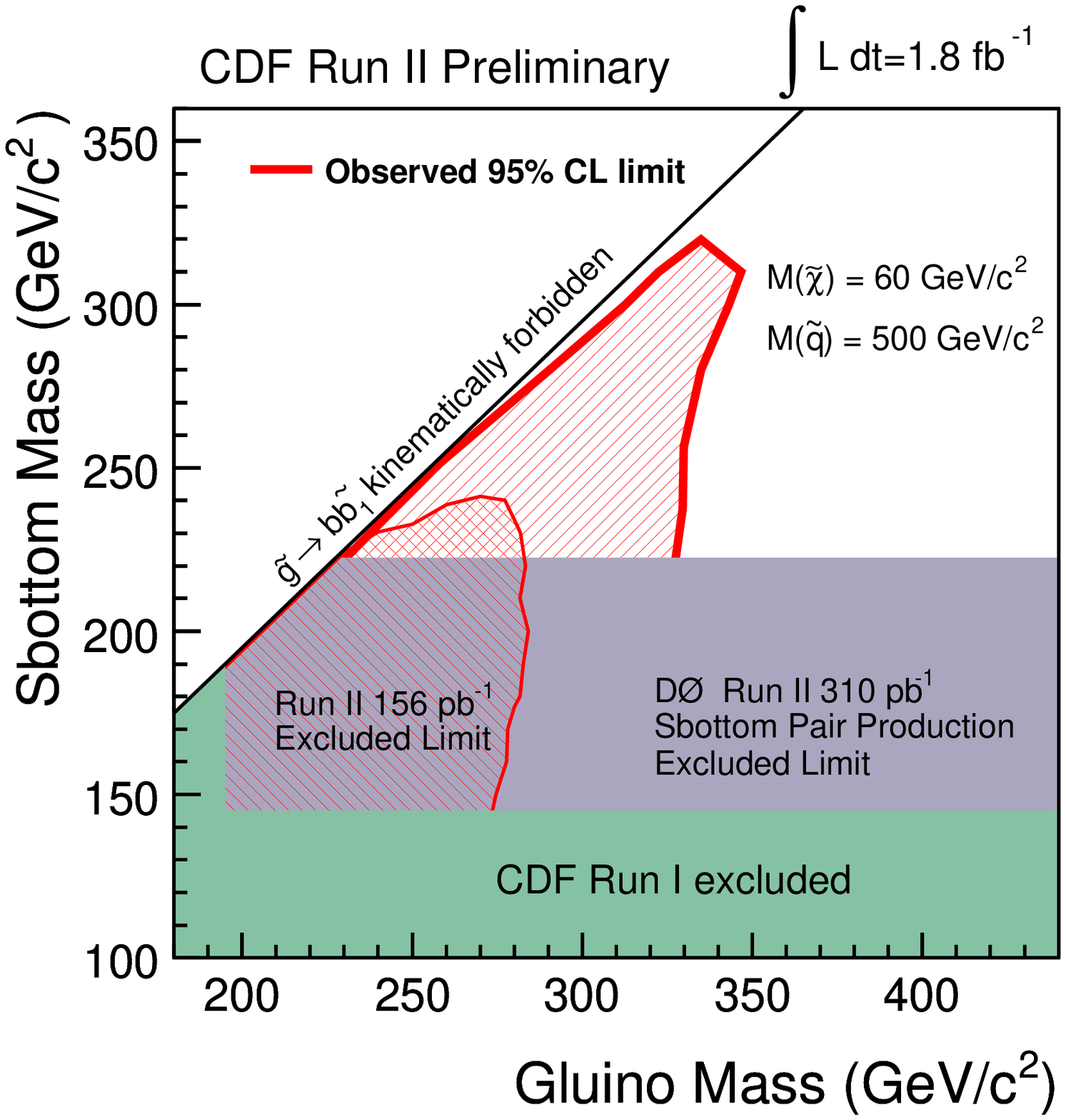}}}
\vspace{-0.05in}
\end{minipage}\hspace*{0.4in}
\begin{minipage}{0.45\textwidth}
\centerline{\mbox{\includegraphics[width=90mm,height=80mm]{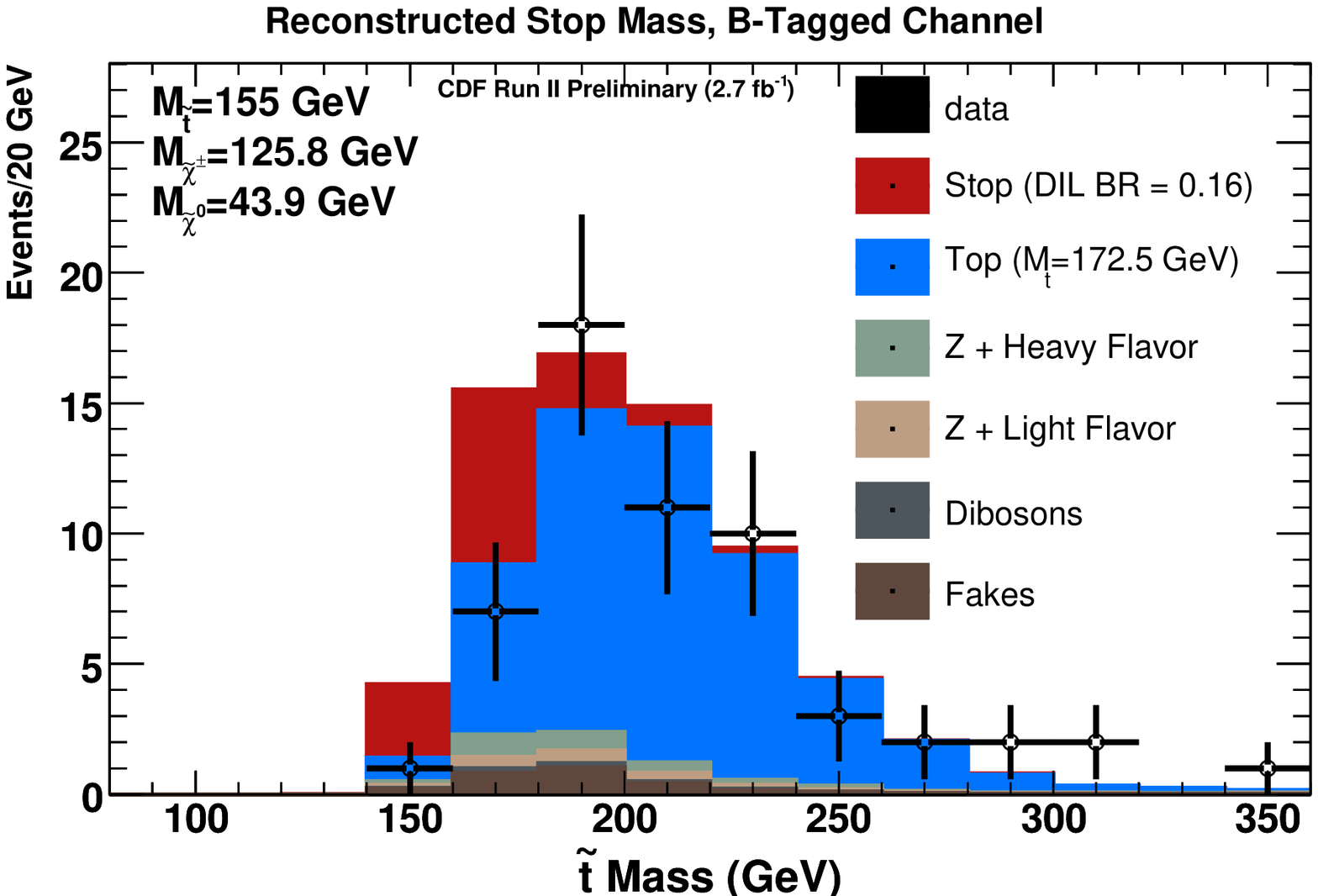}}}
\vspace{-0.04in}
\end{minipage}
}
\caption{{\bf LEFT:} Gluino-sbottom mass plane: 
the brown region shows the area excluded at 95$\%$ C.L. by the search for 
gluino-mediated sbottom production. The dashed black line shows the expected limit. 
Previous exclusion limits are also reported. {\bf RIGHT:} Reconstructed 
stop mass in the b-tagged channel, comparing prediction 
to data. Stop is plotted at the dilepton branching ratio excluded at the 95$\%$ 
level for various chargino, neutralino, and stop masses.} \label{sbottom}

\end{figure*}

In order to further reduce the background contributions and achieve the best
sensitivity to the SUSY signal, a dedicated study has been carried 
out to maximize the signal significance, S/$\sqrt{\rm B}$, within respect to  
relevant observables such as $\ETM$, jet multiplicity and jets transverse energy. 
Two optimization regions are finally defined depending on the mass difference between 
the gluino and the sbottom, $\Delta m$. For small(large) $\Delta m$ optimization, 
19(25) events are observed, compared to a SM background prediction of 
22.0$\pm$3.6(22.7$\pm$4.6). The total systematic uncertainties are 25$\%$(30$\%$)  
for the large(small) $\Delta m$ region, and are dominated by the 3$\%$ uncertainty 
on the jet energy scale and the 5$\%$ uncertainty on the estimation of mistag 
and QCD-multijet background. 
Since no evidence of SUSY signal is found, results are used to set exclusion limits 
using a Bayesian approach at 95$\%$ C.L. The $\gl \gl$ cross section limit is found around 
0.4 pb, nearly independent on the sbottom mass. The exclusion region in the gluino-sbottom 
mass plane is shown in Figure~\ref{sbottom} (left): results from previous analyses 
are extended with the exclusion of sbottom masses up to 
300 GeV/c$^{2}$ for gluino masses below 340 GeV/c$^{2}$. 



\section{SEARCH FOR STOP QUARKS IN TTBAR FINAL STATES}
Due to the large mass of the top quark, the mass splitting in the stop sector can 
be large, allowing $\tilde{t}^{}_{1}$ to be the lightest squark, possibly 
lighter than the top quark. 
In this search, performed using 2.8 fb$^{-1}$ of data~\cite{sqglpublic}, stop pair 
production is searched for in final states mimicking the dilepton plus b-jets 
$t \bar{t}$ signature. Stop quarks are assumed to decay with 100$\%$ BR into 
a b-quark and a chargino. Under the assumption that 
$m^{}_{\tilde{\chi}^{\pm}_{1}}-m^{}_{\tilde{\chi}^{0}_{1}} < m^{}_{W}$, 
the chargino decay into dilepton final states can happen through a variety of channels 
and be significantly enhanced. 
Signal samples are generated for stop masses between 115 and 185 GeV/c$^{2}$ using 
{\sc pythia} and considering the chargino mass between 105.8 and 125.8 GeV/c$^{2}$, 
and the neutralino mass between 43.8 and 88.5 GeV/c$^{2}$. 
The total NLO production cross section of stop pair production  
is estimated using {\sc prospino}. 

The reconstructed mass of stop candidates is used to discriminate stop production from the 
Standard Model background processes. 
Events are required to have two high transverse momentum leptons (electrons or muons, 
p$^{}_{\rm T}>$20 GeV/c), at least one of them isolated.  
Events must have \MET above 20 GeV, and at least two jets with 
E$^{}_{\rm T}>$15(12) GeV in the final state. The Z boson contributions are vetoed using a 
variable based on \MET. Top pair events are suppressed by selecting on a discriminant 
which correlates the azimuthal separations between leptons and the two leading 
jets with the sum of the transverse momenta of all the above objects.  
Two event categories are considered as signal regions: one requires at least one 
b-tagged jet, the other must not contain tagged jets. Slightly different selections 
are used in each case to maximize the sensitivity to the signal. 
The observed number of events in both regions is found 
in good agreement with SM predictions. Figure~\ref{sbottom} (right) shows 
the measured reconstructed mass for $\tilde{t}$ candidate events in the 
b-tagged channel. The expected contribution from stop signal samples is also 
shown assuming a dilepton branching ratio of 0.16. 
Limits at 95$\%$ C.L. are extracted as a function of the dilepton branching ratio, 
for stop masses between 115 and 185 GeV, neutralino masses between 43.9 and 
88.5 GeV, and at chargino masses of 105.8 and 125.8 GeV. 

\section{CONCLUSIONS}
The most recent results on searches for squarks and gluinos 
in events with large missing transverse energy, leptons and multiple jets in 
the final state have been presented. No evidence of New Physics has been 
found yet and stringent exclusion limits have been extracted in several 
MSSM scenarios. With more than 4 fb$^{-1}$ of data already collected, CDF could 
reveal hints of New Physics, or place more severe limits on the 
SUSY parameter space before the start-up of the LHC.  



\begin{thebibliography}{9}
\bibitem{Rp} $\rm R^{}_{P} = (-1)^{3(B-L)+2s}$, where B is the baryon number, L is the lepton number, and s is the spin.
\bibitem{sbottomnew} Most recent updates for this search are available in~\cite{sqglpublic}.
\end{thebibliography}
\end{document}